\documentclass{PoS}
\usepackage{wrapfig,rotating}
\def\GeV{{\rm GeV}}

\title{Theoretical Procedures and the effect on PDFs and
$\alpha_S(M_Z^2)$}

\ShortTitle{Theoretical Procedures and the effect on PDFs and
$\alpha_S(M_Z^2)$}

\author{\speaker{Robert S Thorne}\\
        University College London\\
        E-mail: \email{robert.thorne@ucl.ac.uk}}

\abstract{I consider the effect on partons distribution functions (PDFs)
of changes in the theoretical procedure used in a PDF fit. I consider
using the 3-flavour fixed flavour number scheme instead of the standard
general mass variable flavour number scheme used in the MSTW analysis.
This results in the light quarks increasing at most $x$ values, the gluon
distribution becoming softer at high values of $x$ and larger at small $x$,
and the coupling constant $\alpha_S(M_Z^2)$ falling, particularly at NNLO.
The fit quality also deteriorates.I also consider lowering the kinematic
cut on $W^2$ for DIS data and introducing higher twist terms which are fit
to data. This results in much smaller effects on both PDFs and
$\alpha_S(M_Z^2)$, with changes generally smaller than uncertainties,
except for quarks at very high $x$. I show that the fixed flavour
scheme and variable flavour scheme structure functions differ
significantly for $x \sim 0.01$ at high $Q^2$. I demonstrate that in the
fixed flavour scheme there is a slow convergence of large
logarithmic terms of the form $(\alpha_S\ln(Q^2/m_c^2))^n$ in this regime.
I conclude that some major differences in PDF
sets are largely due to the choice of flavour scheme used.}

\FullConference{XXI International Workshop on Deep-Inelastic Scattering and Related Subject -DIS2013,\\
		22-26 April 2013\\
		Marseilles,France}

\begin{document}

There have recently been improvements in the PDF determinations by 
the various groups, generally making the predictions more consistent. 
However, there are still some large differences which are sometimes much bigger 
than the individual PDF uncertainties \cite{Watt:2011kp,
Forte:2013wc,Ball:2012wy}. This is particularly the case for 
cross sections depending on the high-$x$ gluon.In this article I investigate 
potential reasons, based on different theoretical procedures that can be 
chosen for a PDF fit. 

I first examine the number of active quark flavour used in the 
calculation of structure functions, where there are  two choices 
for how one treats the charm and bottom quark
contributions. 
In the $n_f=3$ Fixed Flavour Number Scheme (FFNS)
$F(x,Q^2)=C^{FF, n_f}_k(Q^2/m_H^2)\otimes f^{n_f}_k(Q^2)$, 
i.e. for $Q^2\sim m_c^2$ massive quarks are only  
created in the final state.
This is exact but does not sum all $\alpha_S^n \ln^n Q^2/m_c^2$ terms in the 
perturbative expansion. The FFNS is known at NLO \cite{Laenen:1992zk}
but not fully at NNLO (${\cal O}(\alpha_S^3C^{FF,3})$).
Approximate results can be derived e.g. \cite{Kawamura:2012cr}, 
(and are sometimes used in fits, e.g. \cite{Alekhin:2012ig})      
but these NNLO corrections are not large except near 
threshold and very low $x$. In a  
variable flavour scheme  one uses the fact that 
at $Q^2 \gg m_c^2$ 
the heavy quarks behave like massless partons and the   
$\ln(Q^2/m_c^2)$ terms are summed via 
evolution. PDFs in different number regions are related 
perturbatively, $f^{n_f+1}_j(Q^2)= A_{jk}(Q^2/m_H^2)
\otimes f^{n_f}_k(Q^2)$ where the
perturbative matrix elements $A_{jk}(Q^2/m_H^2)$
are known exactly to NLO \cite{Buza:1996wv}. 
The original Zero Mass Variable Flavour 
Number Scheme (ZM-VFNS)
ignores ${\cal O}(m_c^2/Q^2)$ corrections in cross sections, i.e. 
$F(x,Q^2) = C^{ZM,n_f}_j\otimes f^{n_f}_j(Q^2)$,
but this is 
an approximation at low $Q^2$ and PDF groups use
a  General-Mass Variable Flavour Number Scheme 
(GM-VFNS) taking one from the two well-defined limits of $Q^2\leq m_c^2$ and 
$Q^2\gg m_c^2$ instead. Some variants are reviewed in \cite{Binoth:2010ra}.

\begin{figure}[htb]
\vspace{-0.8cm}
\centerline{\hspace{-0.7cm}
\includegraphics[width=0.45\textwidth]{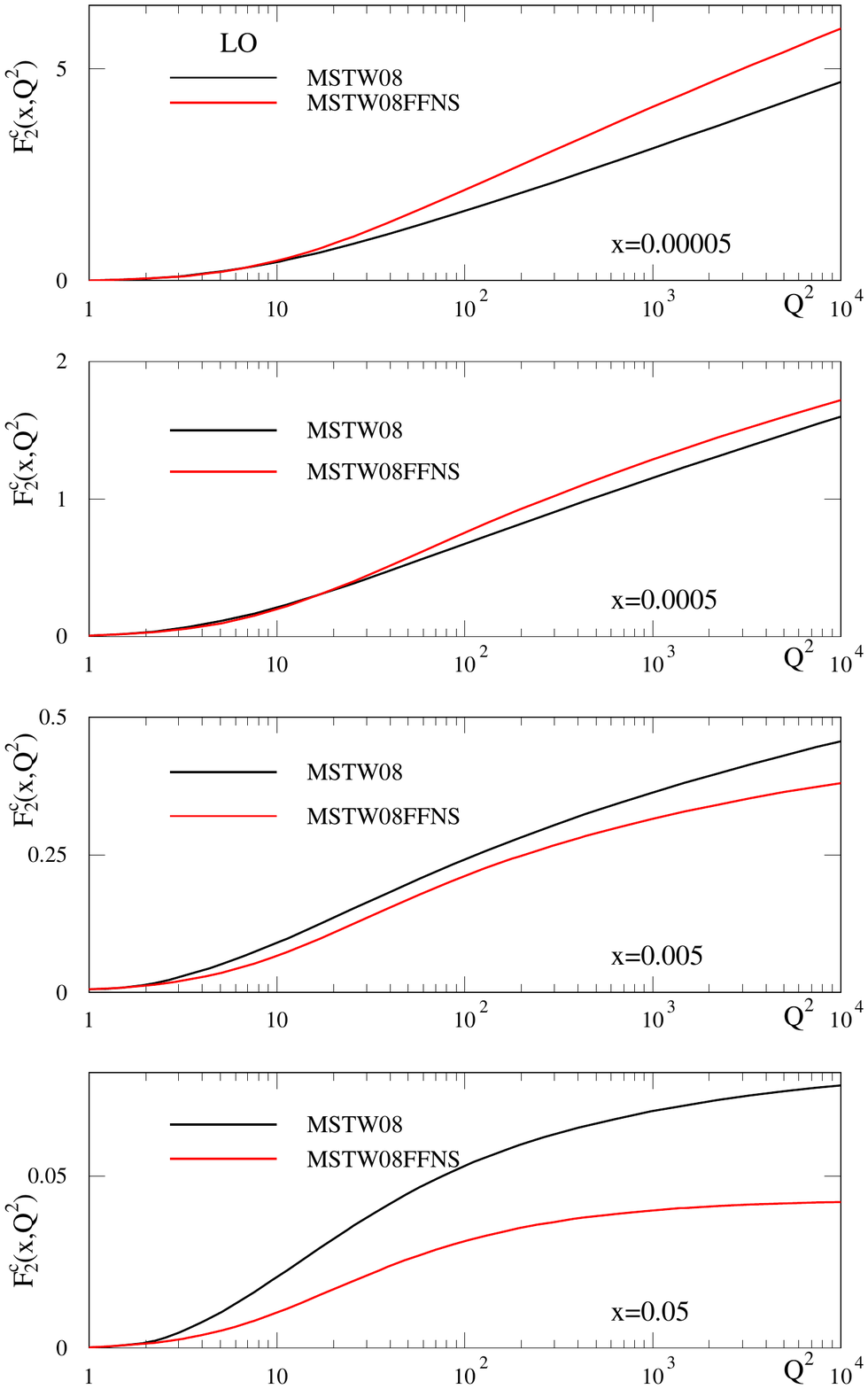}
\hspace{-1.8cm}\includegraphics[width=0.45\textwidth]{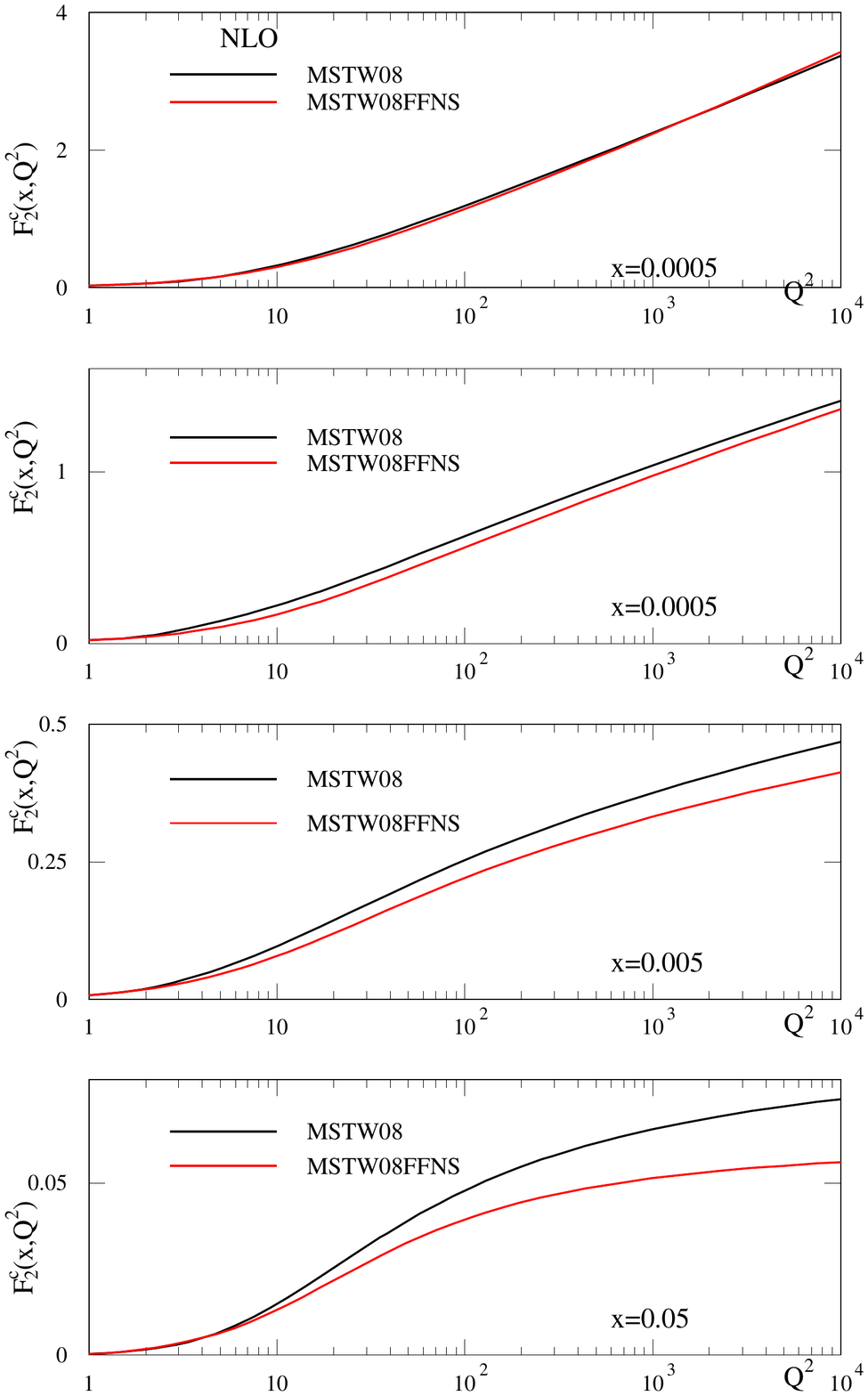}
\hspace{-1.8cm}\includegraphics[width=0.45\textwidth]{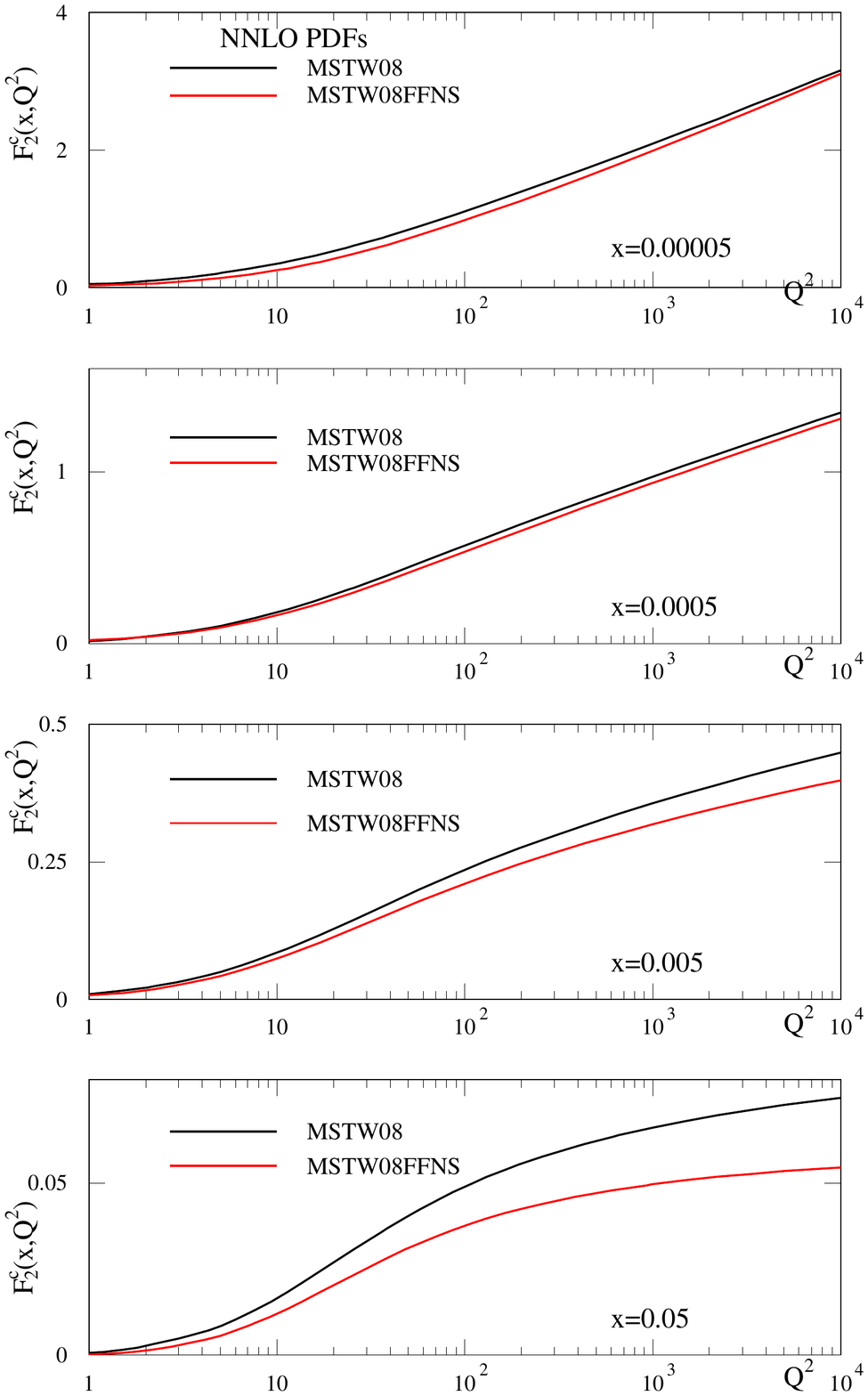}
\hspace{-1.5cm}}
\vspace{-0.9cm}
\caption{$F_2^c(x,Q^2)$ using the FFNS and GM-VFNS at LO, NLO and NNLO.}
\vspace{-0.4cm}
\label{Fig1} 
\end{figure}

The predictions using FFNS and the TR' GM-VFNS \cite{Thorne:2006qt}
for $F_2^c(x,Q^2)$ using the MSTW2008 input distributions 
\cite{Martin:2009iq} are shown in Fig.~\ref{Fig1}.
At LO there is a big difference between the two. At NLO  
$F_2^c(x,Q^2)$ at high $Q^2$ for the FFNS is nearly always lower than for the GM-VFNS, 
significantly so at higher $x\sim 0.01$. For FFNS at 
NNLO only NLO coefficient functions are used, but (various choices of)   
approximate ${\cal O}(\alpha_S^3)$ corrections
give only only minor
increases. There is no dramatic improvement in the agreement between 
FFNS and GM-VFNS at NNLO compared to NLO, contrary to what one might expect.

\begin{wrapfigure}{r}{0.36\columnwidth}
\vspace{-1.3cm}
\centerline{\includegraphics[width=0.5\textwidth]{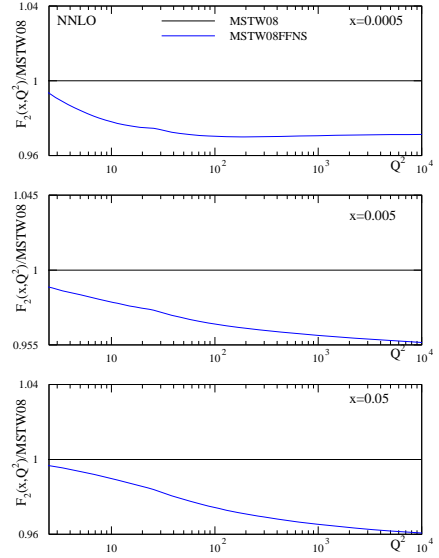}}
\vspace{-1.6cm}
\caption{The ratio of $F(x,Q^2)$ using the FFNS to that using the GM-VFNS.}
\vspace{-0.7cm}
\label{Fig2} 
\end{wrapfigure}

\begin{wrapfigure}{r}{0.67\columnwidth}
\vspace{-1.9cm}
\centerline{\includegraphics[width=0.44\textwidth]{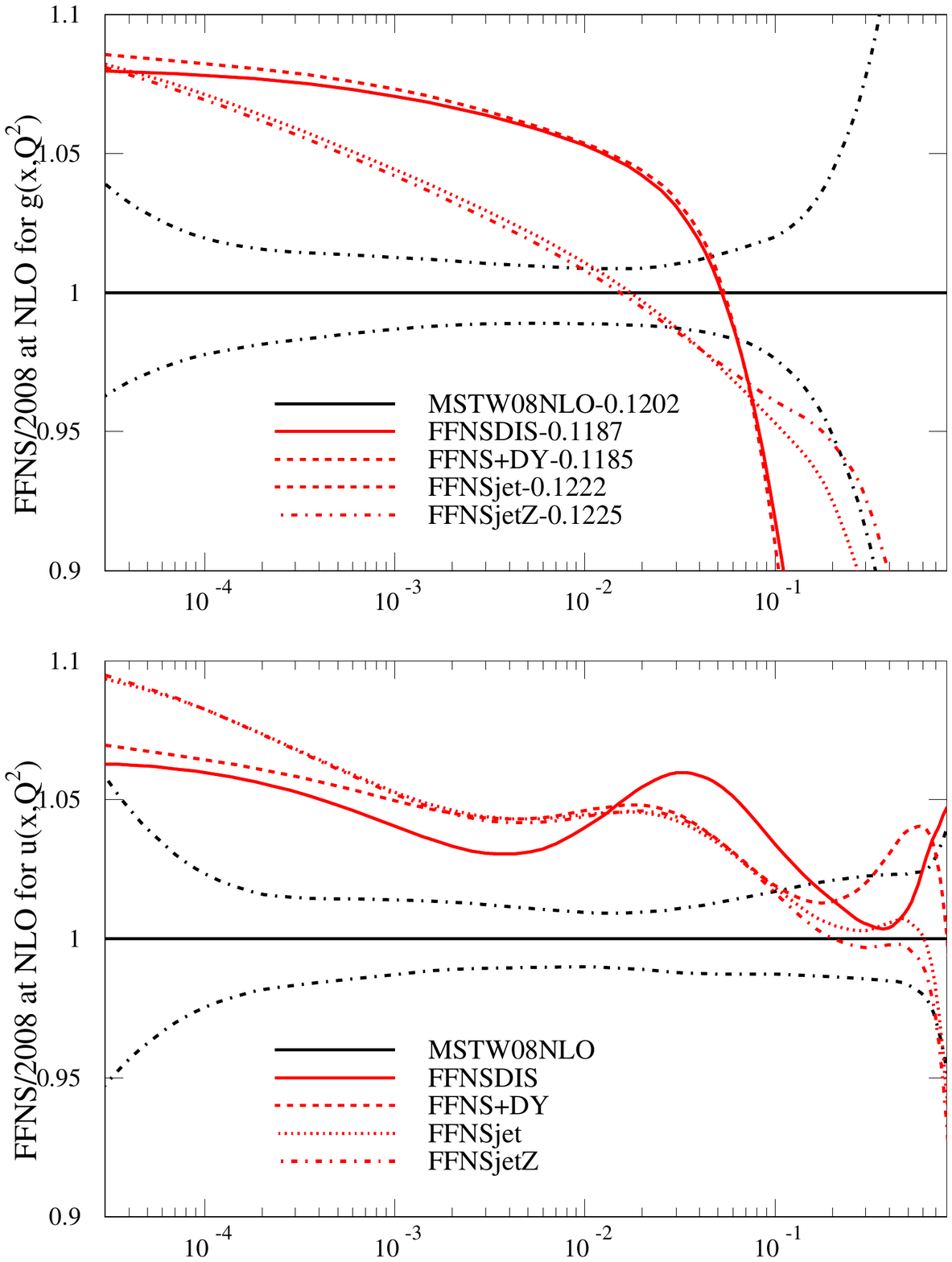}
\hspace{-1.5cm}\includegraphics[width=0.44\textwidth]{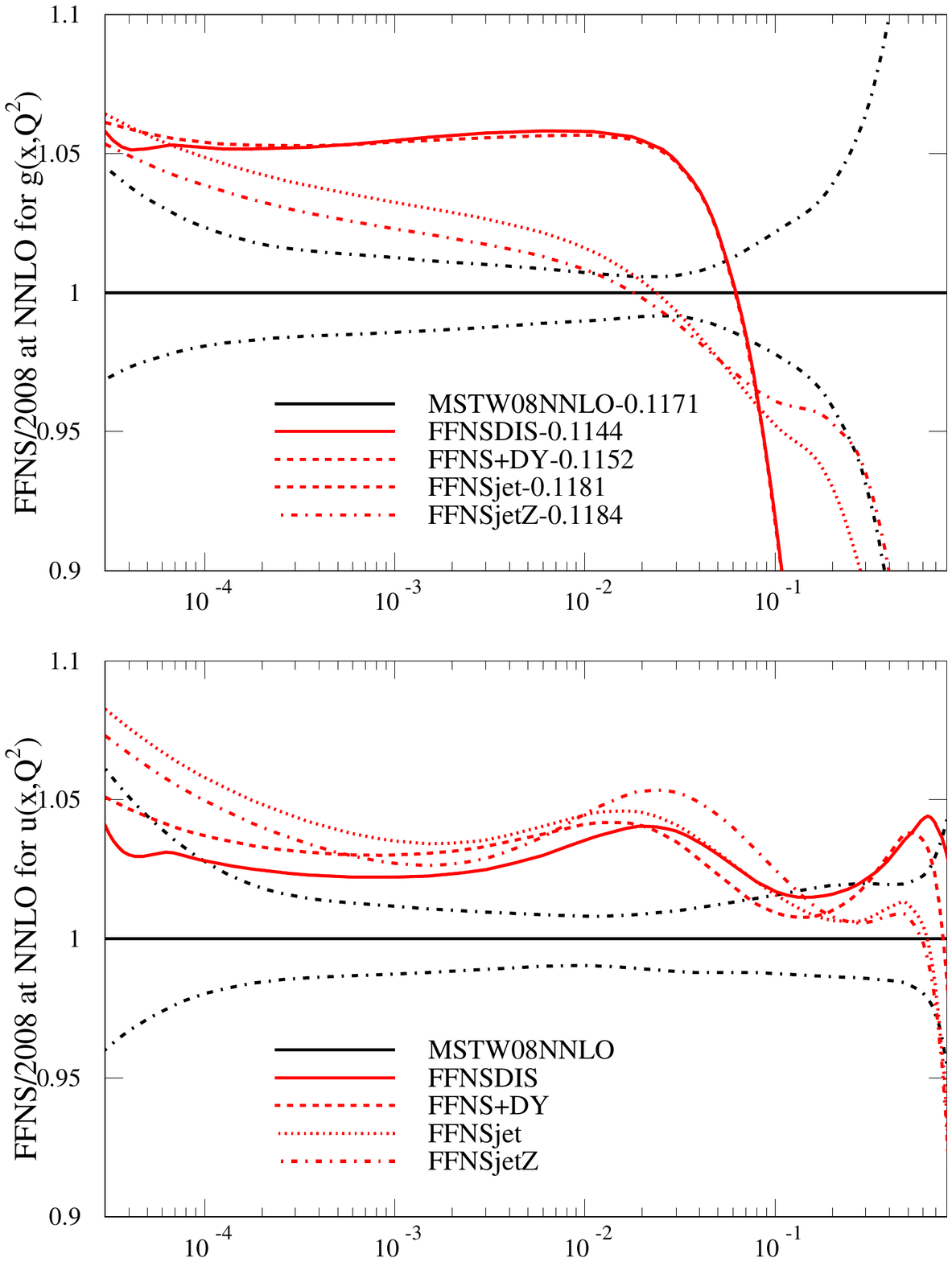}}
\vspace{-1.4cm}
\caption{Ratios of PDFs in various FFNS fits to the MSTW2008 PDFs.}
\vspace{-0.5cm}
\label{Fig3} 
\end{wrapfigure}

This $20$-$40\%$ difference in $F_2^c(x,Q^2)$
can lead to over $4\%$ changes in the total 
$F_2(x,Q^2)$, see Fig~\ref{Fig2}. 
At $x\sim 0.01$ this is mainly due to $F_2^{c}(x,Q^2)$. 
At lower $x$ there is a contribution to the difference from 
light quarks evolving slightly more slowly in FFNS.   
For $x>0.1$ the FFNS and GM-VFNS are very similar. 
In order to test the importance of 
this difference I have extended an investigation in 
\cite{Thorne:2012az} and performed fits using
the FFNS scheme. At NNLO ${\cal O}(\alpha_S^2)$
heavy flavour coefficient functions are 
used as default (which has been done in 
other fits, e.g. \cite{Alekhin:2009ni}). Approximate 
${\cal O}(\alpha_S^3)$ expressions change the 
results very little. Fits are primarily to only DIS and Drell-Yan 
data, but are also extended to Tevatron jet and Drell-Yan 
data using the 5-flavour scheme in these cases. 
The data chosen are as in \cite{Martin:2009iq}. 
The fit quality for DIS and Drell-Yan 
data are at least a few tens of units higher in $\chi^2$ in the FFNS fit 
than in the MSTW2008 fit. 
FFNS is often slightly better for $F_2^c(x,Q^2)$, but is 
flatter in $Q^2$ for $x \sim 0.01$ for the total $F_2(x,Q^2)$. 
When using the FFNS 
the fit quality for DIS and Drell Yan deteriorates by
$\sim 50$ units when Tevatron jet and $W,Z$ 
data are included, as opposed to $10$ units or less when using a GM-VFNS.   
The resulting PDFs evolved up to $Q^2=10,000\GeV^2$ (using variable 
flavour evolution for consistent comparison) are 
shown in Fig.~\ref{Fig3}. The PDFs and $\alpha_S(M_Z^2)$ 
are different in form to 
the MSTW2008 PDFs, with larger light quarks, a gluon which is bigger at 
low $x$ and much smaller at high $x$ and a smaller $\alpha_S(M_Z^2)$. 
Some similar differences have 
been noted in \cite{CooperSarkar:2007ny,Ball:2013gsa}.
Using FFNS rather than GM-VFNS leads to much larger changes than any
variation in choice of GM-VFNS \cite{Thorne:2012az}.

\begin{wrapfigure}{r}{0.68\columnwidth}
\vspace{-1.3cm}
\centerline{\includegraphics[width=0.44\textwidth]{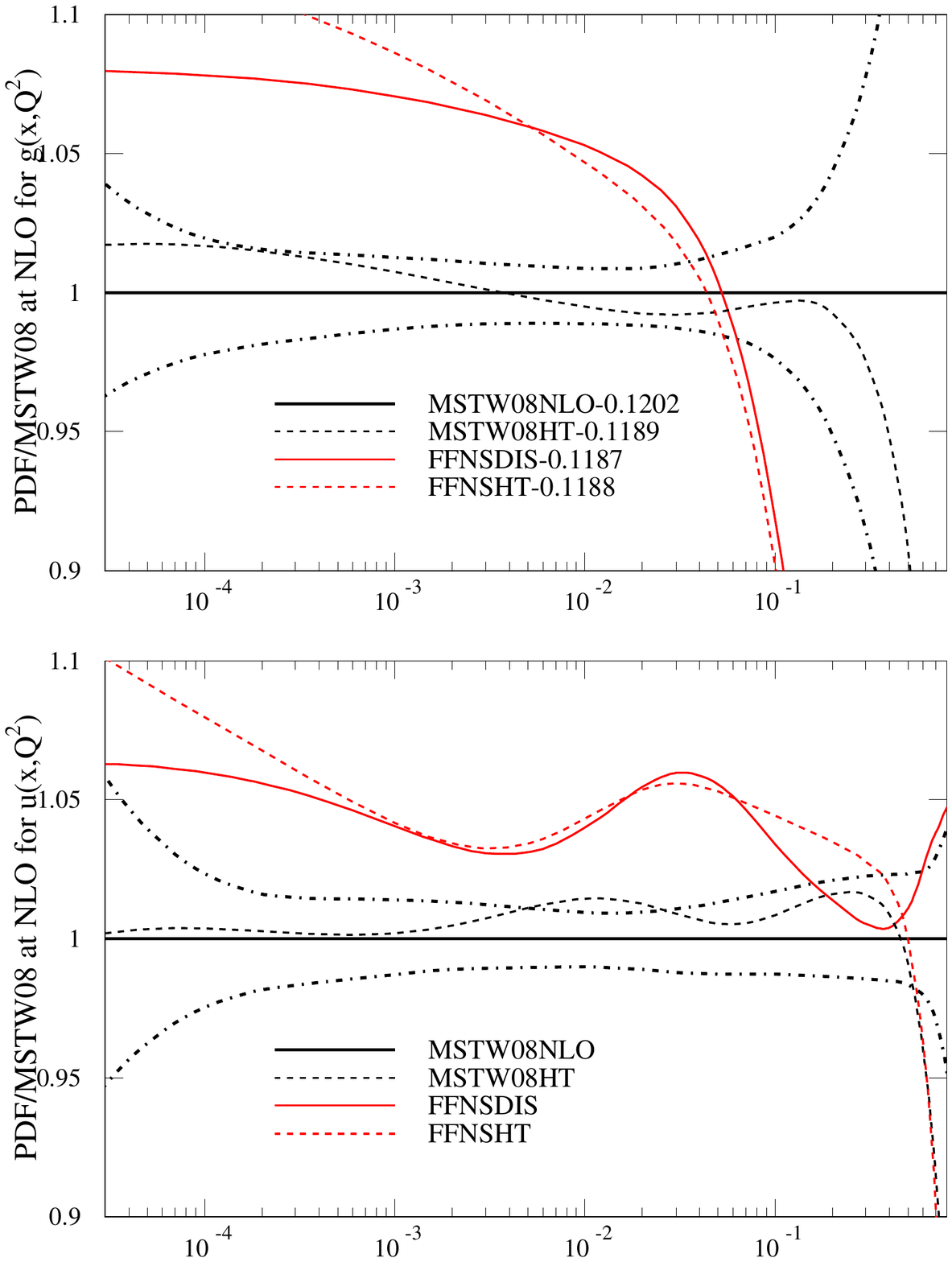}
\hspace{-1.5cm}\includegraphics[width=0.44\textwidth]{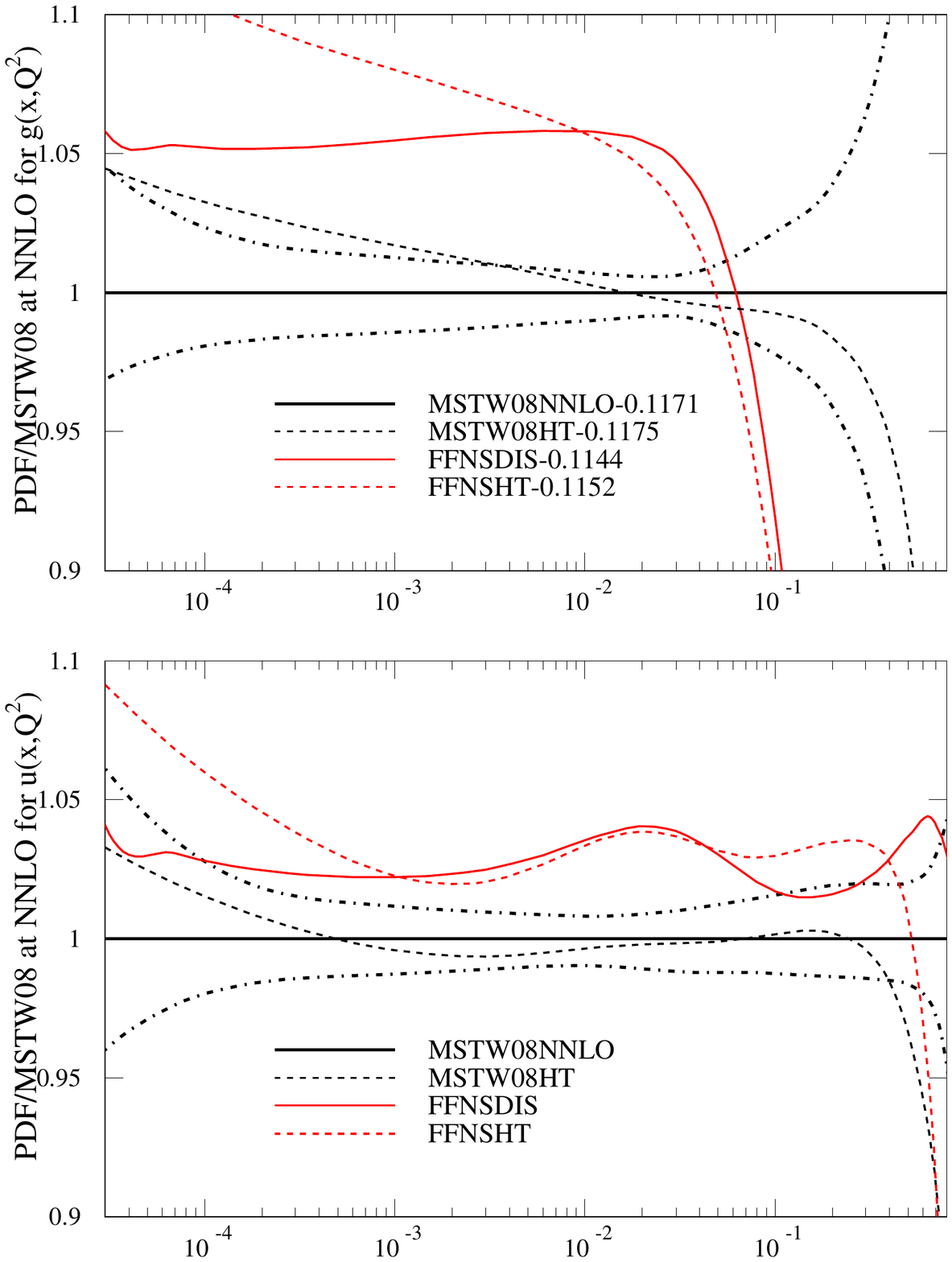}}
\vspace{-1.6cm}
\caption{Ratios of PDFs with higher twist corrections to PDFs without.}
\vspace{-0.6cm}
\label{Fig4} 
\end{wrapfigure}

I have also investigated the effect of lowering the $W^2$ cut to 
$5~\GeV^2$ and parameterising higher twist corrections 
in the form $(D_i/Q^2)F_2(x,Q^2)$ in 13 bins of $x$, and fitting the $D_i$ 
and PDFs simultaneous, as in \cite{Martin:2003sk}. The $D_i$ are 
similar to this older study, though larger at the smallest $x$. 
The effect on the PDFs and $\alpha_S(M_Z^2)$ is small, 
using either FFNS or GM-VFNS, except for very high-$x$ quarks, 
as shown in Fig.~\ref{Fig4}. I also
perform FFNS fits restricting higher twist from the lowest $x$ values 
and omitting the less theoretically clean nuclear target 
data (except dimuon cross sections, which constrain the strange quark). 
This results in values of $\alpha_S$ of
$\alpha_S(M_Z^2)=0.1179$ at NLO and $\alpha_S(M_Z^2)=0.1136$ at NNLO, 
very close to those in \cite{Alekhin:2009ni}, where the scheme choice, data 
types, and form of higher twist  
(and the resulting PDFs) are similar.

\begin{wrapfigure}{r}{0.68\columnwidth}
\vspace{-1.6cm}
\centerline{\includegraphics[width=0.44\textwidth]{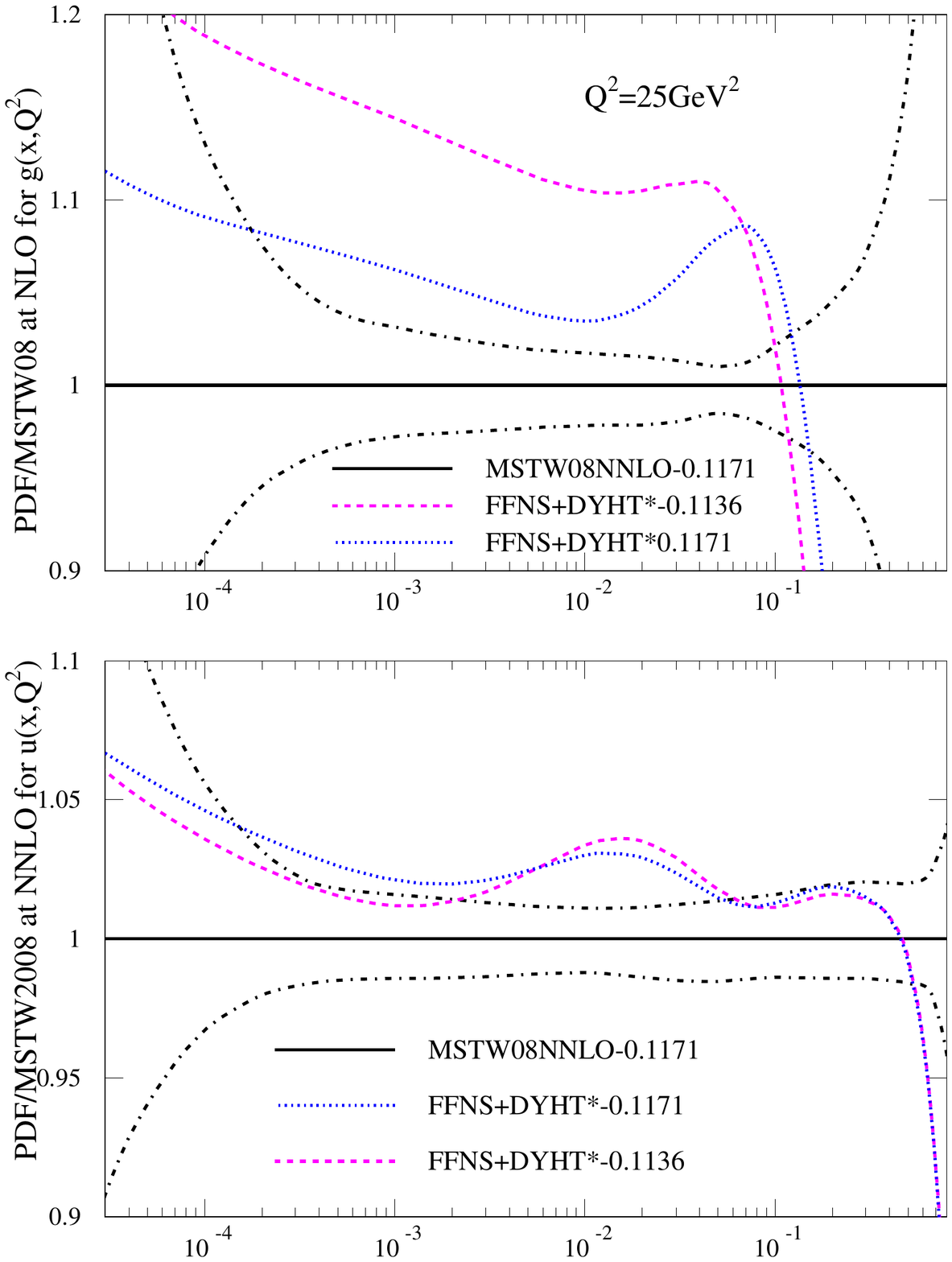}
\hspace{-1.5cm}\includegraphics[width=0.44\textwidth]{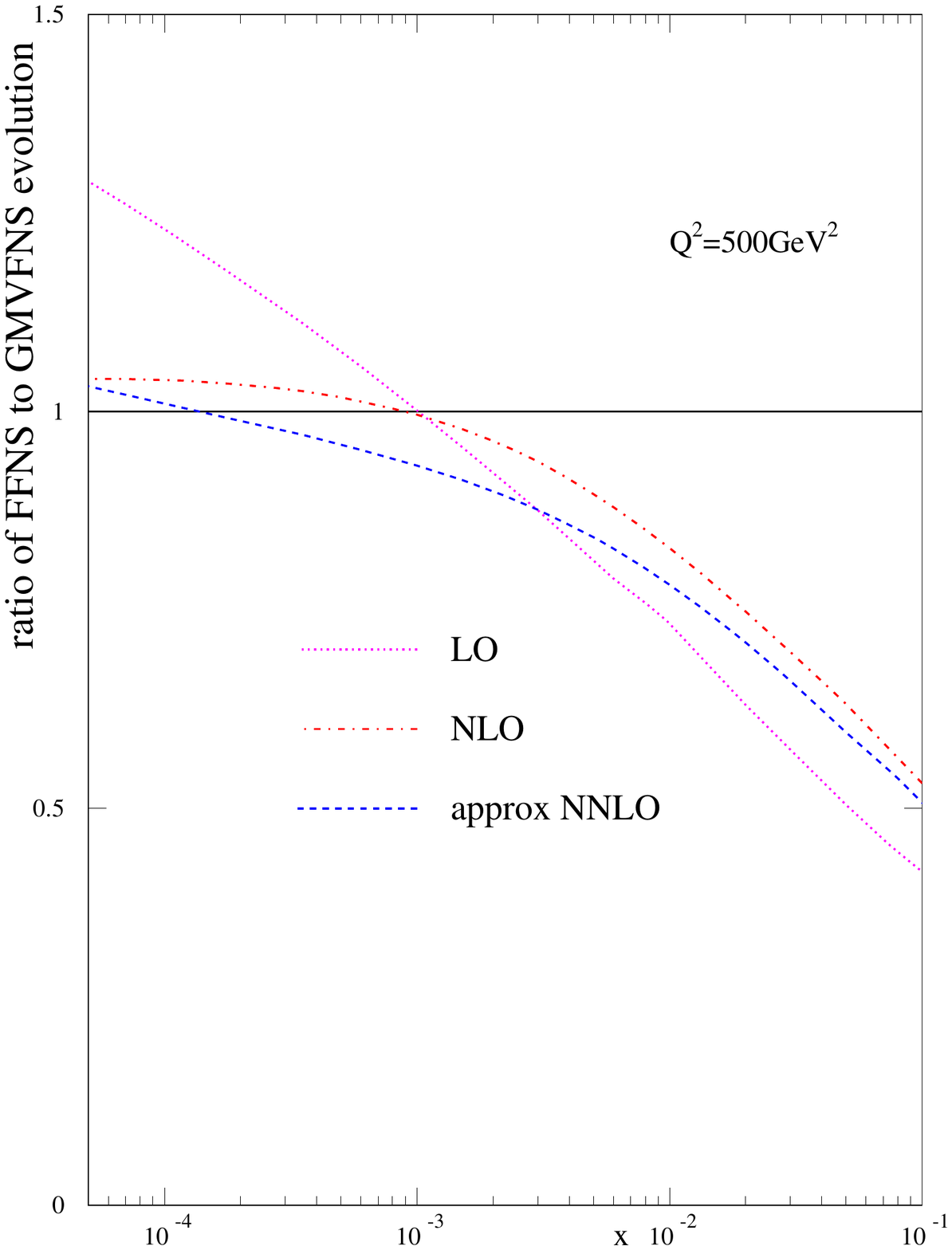}}
\vspace{-1.5cm}
\caption{The ratio of FFNS fits with both free (red) and fixed $\alpha_S(M_Z^2)$ (blue)
to the MSTW2008 PDFs (left) and the ratio of $dF_2^c/d\ln Q^2$ using
the FFNS to that using the GM-VFNS (right).}
\vspace{-0.4cm}
\label{Fig5} 
\end{wrapfigure}

I also perform fits where $\alpha_S(M_Z^2)$ 
is fixed to the higher value obtained in the GM-VFNS. This results in the 
FFNS gluon being a little closer to that using GM-VFNS, as shown in the 
left of  Fig.~\ref{Fig5}
and very similar to \cite{Ball:2013gsa}, where studies are performed with 
fixed $\alpha_S(M_Z^2)$. The fit quality to DIS and low-energy DY data in
the FFNS fit is 8 units worse when $\alpha_S(M_Z^2)=0.1171$
than for 0.1136. The fit to HERA data is better, but worse for fixed  
target data. 
One can understand the need for $\alpha_S$ to be smaller in 
FFNS. To compensate for

\noindent  
 smaller $F_2^c(x,Q^2)$ at 
$x \sim 0.05$ the FFNS gluon must be bigger in this region, and 
from the momentum sum rule, is smaller at 
high $x$.The correlation between the high-$x$ gluon and 
$\alpha_S$ when fitting high-$x$ DIS data drives $\alpha_S$ down
(for reduced gluon the quarks fall with $Q^2$ too quickly, 
hence the need to  lower $\alpha_S$),
requiring  the small $x$ gluon to even bigger, until stability 
is reached.

%\begin{wrapfigure}{r}{0.8\columnwidth}
%\vspace{-0.3cm}
%\centerline{\includegraphics[width=0.4\textwidth]{ffnspartcompnlo}
%\includegraphics[width=0.4\textwidth]{ffnspartcompnnlo}}
%\vspace{-0.5cm}
%\caption{xxxxxxxxxxxx.}
%\label{partratios} 
%\end{wrapfigure}

To explain the differences between FFNS and GM-VFNS evolution,
shown for $Q^2=500\GeV^2$ in the right of Fig.~\ref{Fig5}, 
I consider high $Q^2$. At LO in the FFNS (setting all scales as $Q^2$)
\begin{eqnarray}
 F_2^{c,1,FF} & = & 
\alpha_S\ln(Q^2/m_c^2) p^0_{qg}\otimes g 
+ {\cal O}(\alpha_S \cdot g) \equiv \alpha_S A_{Hg}^{1,1}
\otimes g + {\cal O}(\alpha_S \cdot g) ,\cr
\to  d\,F_2^{c,1,FF}/d \ln Q^2 & = & \alpha_S p^0_{qg}\otimes g
 + \ln(Q^2/m_c^2)d \, (\alpha_S p^0_{qg}\otimes g)/d\ln Q^2 + \cdots.
\end{eqnarray}
At LO in the GM-VFNS, where $F_2^{c,1,VF}=(c + \bar c) = c^+$ we have
\begin{equation}
d\,c^+/d \ln Q^2 = \alpha_S \,p^0_{qg}\otimes g  +
\alpha_S\, p^0_{qq} \otimes c^+, \quad 
c^+ \equiv \alpha_S\ln(Q^2/m_c^2) p^0_{qg}\otimes g + \cdots \equiv \alpha_S A_{Hg}^{1,1}\otimes g + \cdots 
\end{equation} 
%so the second term is formally ${\cal O}(\alpha_S^2 \ln(Q^2/m_c^2))$.  
The first terms in each expression 
%is of the form $\alpha_S\ln(Q^2/m_c^2)$ and 
are equivalent. 
The difference between the LO expressions is
\begin{equation}
d(F_2^{c,1,VF}\!\!\!\!-F_2^{c,1,FF})/d \ln Q^2 = \alpha_S^2 \ln(Q^2/m_c^2)p^0_{qg}
\otimes (p^0_{qq} +\beta_0 -p^0_{gg}) \otimes g + \!\cdots
\equiv P^{\rm LO}_{VF-FF} \otimes g + \!\cdots
\end{equation}  
where $\beta_0 =9/(4\pi)$. The effect of $p^0_{gg}$
is negative at high $x$ and positive at small $x$. That of 
$p^0_{qq}$ is negative at high $x$, but smaller than 
$p^0_{gg}$. 
Hence, the difference is large and positive at high $x$ and 
large and negative at small $x$, as observed in Fig~\ref{Fig5}. 
Moreover, this difference must be eliminated at NLO by 
defining the leading-log term in the FFNS expression
to provide cancellation, i.e. 
\begin{equation}
F_2^{c,2,FF}\!\!= \alpha^2_S A_{Hg}^{2,2} \otimes g + \cdots = 1/2
\alpha^2_S \ln^2(Q^2/m_c^2) p^0_{qg}\otimes (p^0_{qq} +\beta_0 -p^0_{gg}) \otimes g   + {\cal O}(\alpha^2_S \ln(Q^2/m_c^2)).
\end{equation}
up to quark mixing corrections and sub-dominant terms. 
In the NLO evolution all  
${\cal O}(\alpha_S^2\ln(Q^2/m_c^2))$ terms cancel in the difference. 
However, the derivative of $F_2^{c,2,FF}$ contains 
$1/2
\ln^2(Q^2/m_c^2) d\, \bigl(\alpha^2_S p^0_{qg}\otimes (p^0_{qq} +\beta_0 -p^0_{gg}) \otimes g\bigr)/d\, \ln Q^2$
which does not cancel. This leads to
$P^{\rm NLO}_{VF-FF} = 1/2 \alpha_S \ln(Q^2/m_c^2)(p^0_{qq} +2\beta_0 -p^0_{gg}) 
\otimes P^{\rm LO}_{VF-FF}$.   
The factor of $(p^0_{qq} +2\beta_0 -p^0_{gg})$ is
large, positive at high $x$ and negative at small $x$, but 
not until smaller $x$ than at LO. Therefore, $P^{\rm NLO}_{VF-FF}$
is large and positive at high $x$, negative for 
smaller $x$ and positive for extremely small $x$. This explains 
the behaviour correctly.
Repeating the argument at NNLO  
$P^{\rm NNLO}_{VF-FF}=1/3 \alpha_S \ln(Q^2/m_c^2)(p^0_{qq} +3\beta_0 -p^0_{gg})\otimes
P^{\rm LO}_{VF-FF}$ 
This is large and positive at high $x$ then changes sign 
twice but stays small until becoming negative at tiny
$x$. Again this explains the behaviour correctly.

\begin{wrapfigure}{r}{0.46\columnwidth}
\vspace{-0.8cm}
\centerline{\includegraphics[width=0.5\textwidth]{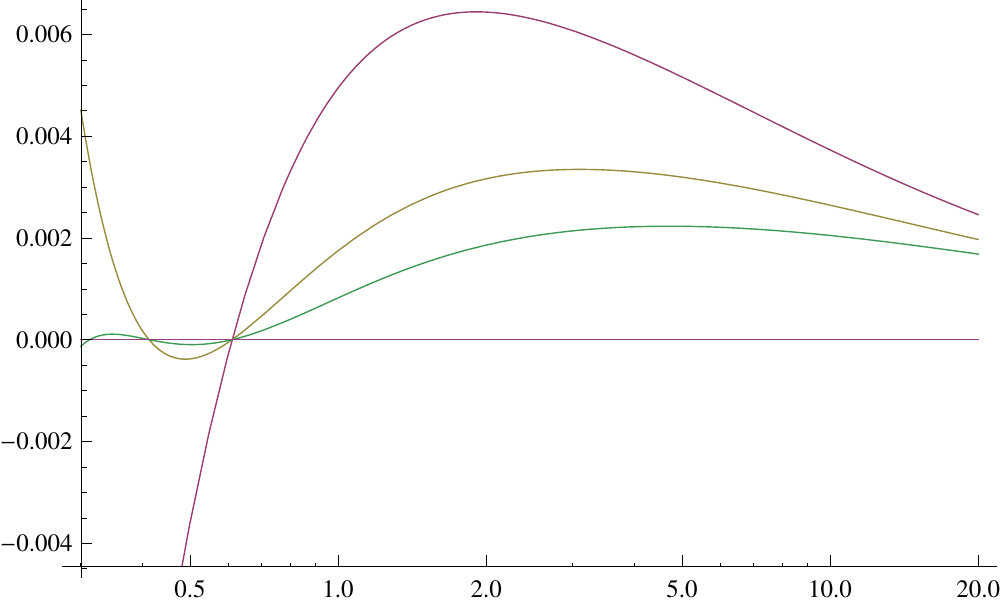}}
\vspace{-0.5cm}
\caption{The effective anomalous dimension $\gamma_{VF-FF}(N)$ at LO (purple), 
NLO (brown) and NNLO (green).}
\vspace{-0.3cm}
\label{Fig6} 
\end{wrapfigure}

To look at the effect of this dominant high-$Q^2$ difference 
between GM-VFNS and FFNS evolution, it is 
useful to define the moment space anomalous dimension 
$\gamma_{VF-FF}$ obtained from $P_{VF-FF}$.
This is shown at LO, NLO 
and NNLO for $Q^2=500\GeV^2$ in Fig.~\ref{Fig6}. At high $Q^2$, values of 
$x\sim 0.05$ correspond to $N\sim 2$, where  $\gamma_{VF-FF}$ only tends to 
zero slowly as the perturbative order increases. This explains why FFNS 
evolution for $x \sim 0.05$ only slowly converges to the GM-VFNS result
with increasing order.

\section*{Acknowledgements}

I would like to thank A. D. Martin, W. J. Stirling  and G. Watt for
numerous discussions on PDFs. This work is
supported partly by the London Centre for Terauniverse Studies (LCTS),
using funding
from the European Research Council via the Advanced Investigator Grant 267352.

\end{document}